\documentclass{appolb}
\usepackage{graphicx}
\usepackage{amssymb,amsmath,amsfonts}

\newcommand{\eq}{\begin{equation}}
\newcommand{\qe}{\end{equation}}
\newcommand{\eqa}{\begin{eqnarray}}
\newcommand{\qea}{\end{eqnarray}}

\newcommand{\QCD}{\mbox{\tiny{QCD}}}
\newcommand{\lqcde}[1]{\Lambda_{\QCD}^{#1}}
\newcommand{\lqcd}{\Lambda_{\QCD}}

\newcommand{\de}{\mathrm{d}}

\begin{document}

\title{Quarkyonic percolation in dense nuclear matter
\thanks{Presented at Excited QCD 2012, May 6--12, Peniche, Portugal.}
}
\author{Stefano Lottini\footnote{Speaker.}
\address{ITP, J.W.~Goethe Universit\"at, Max-von-Laue-Stra\ss{}e 1, \\ 60438 Frankfurt am Main, Germany}
\\
\vspace{0.35cm}{Giorgio Torrieri
}
\address{FIAS, J.W.~Goethe Universit\"at, Ruth-Moufang-Stra\ss{}e 1, \\ 60438 Frankfurt am Main, Germany}
}
\maketitle
\begin{abstract}
		We examine the phase diagram of hadronic matter when the                      
		number of colours $N_c$, as well as temperature and density, are varied.      
		We show that in this regime a new percolation phase transition is             
		possible, and examine the implications of this transition for                 
		extrapolations to physical QCD of the large-$N_c$ limit.                      
\end{abstract}

\section{Introduction}
A powerful tool to investigate the features of non-perturbative QCD is the large-$N_c$ limit \cite{largenc}:
if the number of colours is sent to infinity in a controlled way, the theory undergoes huge simplifications
and a number of results can be easily obtained. The underlying assumption that $N_c=3$ is ``almost infinity'',
however, seems to fail when it comes to predicting nuclear properties (as opposed, for instance, to hadronic ones).

In the following we build a semiclassical model, based on percolation theory, to describe the possibility of a hidden 
phase transition along $N_c$: its ingredients, the shape of a baryon and the likelihood for two quarks to perform some ``exchange'',
can lead to a percolating system, meaning that, with a high enough density, we can have only local (i.e.~confined) interactions
yet achieve a global effect of correlations (exchange of quantum numbers) on a large scale.

The final output of the model, a line separating ordinary confinement from a ``confined percolating'' phase, will then be
compared to the deconfinement curves to assess the existence of a region, in the $T$-$\mu$-$N_c$ phase space,
where this regime is physically realised.
A more thorough discussion on this model can be found in \cite{percdensity}.

\section{The percolation model}
\label{sec:percmodel}
We model dense baryonic matter by a regular cubic lattice with baryons sitting on the sites;
each baryon will contain $N_c$ quarks, independently located according to a hard-sphere distribution
with radius $\lqcde{-1}$:
\eq
	f(\mathbf{x}) \propto \Theta(1 - \lqcd|\mathbf{x}-\mathbf{x}^\mathrm{center}|)\;\;.
\qe
We parametrize density via the ratio $\epsilon$ between the lattice spacing and twice the spheres' radius, so that,
with $\overline{\rho}_0 = \lqcde{3}/8$ (the reference case of \cite{percprl} with baryons exactly touching each other),
we now have $\overline{\rho} = \overline{\rho}_0 \epsilon^{-3}$.
Thinking of the large-$N_c$ limit (where baryons are felt as quasi-static),
we argue this description does not introduce large systematics at least down to
$N_c \sim \mathcal{O}(10)$, the coordination number of the 3D lattice considered \cite{vanderwaals}.

The other important ingredient of the model is how quarks communicate with each other; this is encoded in
a ``squared propagator'' function $F(y)$, the probability of two quarks at distance $y$ to exchange energy/momentum.
As long as $F$ scales as $\lambda/N_c$ and has a sharp drop around a distance $r_T/\lqcd$ ($r_T\sim 1$),
we expect the result to be roughly independent of its exact shape; we consider three functions, corresponding to
a step-function in coordinate space, a step-function in momentum space, and an exponentially screened version of the 
latter:\footnote{The maxima of $F_K$ beyond the first one, albeit tiny, can trigger direct large-distance exchanges that we regard
 as nonphysical: in the following, we will rather consider $F_S$ to get rid of this (high-density) effect.}
\eqa
	F_T(y) &=&  \frac{\lambda}{N_c} \;\Theta\Big(1-\frac{y}{r_T/\lqcd}\Big) \;\;;\\
	F_K(y) &=&  \frac{\lambda}{N_c} \; \frac{2r_T^2}{\pi y^2} \sin^2\Big( \frac{y}{r_T/\lqcd} \Big) \;;
		\;F_S(y) =  F_K(y) e^{-M|y|} \;\;.
\qea

With the baryon description and the ``propagator'' as input, together with the density $\epsilon$, we build the probability
$p$ that two baryons, with centres $\mathbf{x}_{A,B}$, undergo some exchange on a short timescale $\sim \lqcde{-1}$ as:
\eq
	p(N_c) = 1 - \Bigg[
		\int f_A(\mathbf{x}_A)\de^3\mathbf{x}_A
		\int f_B(\mathbf{x}_B)\de^3\mathbf{x}_B
		\Big( 1 - F(|\mathbf{x}_A-\mathbf{x}_B|) \Big)
	\Bigg]^{(N_c)^\alpha};
\qe
in the above formula (evaluated numerically for several choices of
$\lambda, r_T, N_c\in[2$:$80], \epsilon\in[0.8$:$1.4]$ and the three $F$ above),
the quark-quark result is raised to a power that can be either $N_c$ or $N_c^2$, depending on
whether we think of the exchanges as quark ``flips'' or ``interactions'' respectively. It has been seen in \cite{percprl}
that consistency with the large-$N_c$ picture requires the latter choice, hence we set $\alpha=2$ from now on
($\alpha=1$ would give a decreasing $p(N_c)$, see \cite{percdensity} for more discussion).
We remark that the correlations among quarks, neglected in the above equation, were seen in \cite{percprl} to
amount to just a few percent correction.

This work is an extension of \cite{percprl} to a variable-density setting: in particular, when $\epsilon \ll 1$ baryons
start to overlap, with the effect that limiting exchanges to nearest-neighbours would introduce systematic errors.
For that reason, we implement a renormalisation-group-inspired approach, evaluating a set of $p_i$ for nine relative
distances (in lattice units) ranging from $(1,0,0)$ to $(2,2,2)$.
These are then converted, with dedicated Monte Carlo simulations,
into a probability $p_b(p_i)$ for a cell of $b^3$ sites to achieve side-to-side percolation: the wrapping probability
obtained in this way can finally be compared with the expected scaling, known from percolation theory \cite{stauffer},
to extract the threshold value $N_c^*$ at which percolation appears. As a consistency check, we verify substantial independence
of the outcome of the cell size $b$ in the range $3\leq b \leq 7$.

\begin{figure}[htb]
\centerline{
	\includegraphics[width=1.0\textwidth,height=0.25\textwidth]{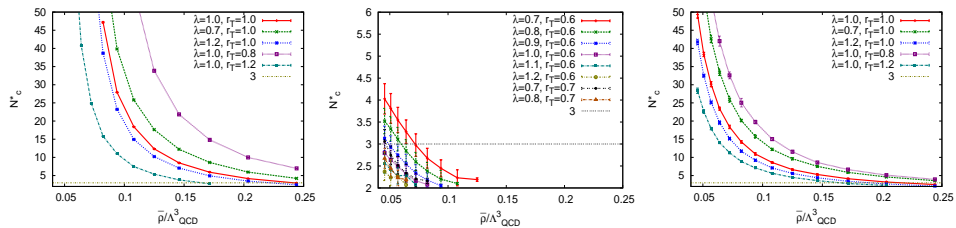}}
	\caption{Plots of $N_c^*$ as a function of the (adimensional) density, for a few sample
		choices of the parameters $\lambda$, $r_T$. Propagators are, from left to right: $F_T$, $F_K$, $F_S$.}
\label{Fig:final_perc_curves}
\end{figure}

The final result is a set of curves (for different propagators and different $\lambda, r_T$)
in the form $N_c^*(\overline{\rho})$ (Fig.~\ref{Fig:final_perc_curves}):
the translation of the density $\overline{\rho}$ into a physical baryonic density $\rho$ raises, in principle, some ambiguities \cite{percdensity},
but in practice the effect is negligible, so in this work they will be treated as coincident.

\section{Deconfinement line in the $T$-$\mu$-$N_c$ space}
The next step is to locate the deconfinement line in the density-$N_c$ plane: above that line, indeed, baryons do not exist
anymore, and the percolation model loses its validity. The deconfinement line, then, is essential
in assessing whether the percolating phase is actually realised: in the following we show how it is obtained, first 
at zero temperature and then at $T>0$.

The densities considered in the previous Section are the order of one baryon per baryonic size (i.e.~$\rho_B\sim \lqcde{3}/8$):
there, according to \cite{quarkyonic}, matter is still confined -- at least for $N_c$ large enough.
A parametric estimate of the zero-temperature deconfinement baryon chemical potential $\mu_0$
can be obtained considering that the quark-hole screening ($\sim \mu_q^2 N_c N_f$, with $N_f$ quark flavours)
has to be the same order as the anti-screening gluon loop ($\sim N_c^2$): providing the correct dimensions with $\lqcd$,
we write\footnote{Unless specified by the subscript ``$q$'', all chemical potentials are baryonic.}
$\mu_0 = N_c^{3/2} N_f^{-1/2} \lqcd$;
neglecting three- (or more) body interactions, this $N_c$-scaling is quite a firm expectation \cite{vanderwaals}.
On the other hand, the $\mu=0$ deconfinement point is $N_c$-independent, phenomenologically described by $T_c\sim (2/3)\lqcd$: we opt for
a ``minimal Ansatz'' to parametrise the deconfinement line in the $T-\mu$ plane as
\eq
	1- \theta^2 = \left(\frac{\mu^{\mathrm{conf}}_B}{\mu_0}\right)^2 \;\;,\;\; \theta = \frac{T}{T_c} \simeq \frac{3}{2}\frac{T}{\lqcd}\;\;.
	\label{eq:theta_param}
\qe
Employing the ideal gas formulas, we can express the relation (so far, at $T\sim 0$)
between (baryonic) density $\rho_B$ and chemical potential $\mu_B$ as
\eq
	\rho_B = \frac{4\pi g_f g_s}{(2\pi)^3} \int_0^{\infty} \Bigg\{
		\frac{p^2~\de p}{\exp\Big[\frac{1}{T}\Big( \sqrt{p^2+m^2}-\mu_B \Big)\Big]+1}
	-\underbrace{[\mu_B\to-\mu_B]}_{(\star)}\Bigg\}
	\;
\qe
(the second term accounting for antibaryons), where $g_f$ is a flavour multiplicity (one for $N_f=1$, $N_f(N_f-1)$ for more),
$g_s$ is a spin-multiplicity factor
($g_s=2$ for spin-$1/2$ baryons), and $m = N_c\lqcd$ is the baryon mass.
With a change of variables designed to put the $N_c$-scaling in evidence (note that we have automatically $\gamma=\sqrt{N_f}$),
\eq
	\gamma = \frac{\sqrt{N_c}}{\mu_0}m  \;\;;\;\;
	\alpha = \frac{\sqrt{N_c}}{\mu_0}p  \;\;;\;\;
	\beta = \frac{\lqcd}{T}  \frac{N_c}{\sqrt{N_f}}\;\;,
\qe
we rewrite the density as
\eq
	\rho_B = \frac{4\pi g_fg_s}{(2\pi)^3} \frac{N_c^3}{N_f^{3/2}}\lqcde{3}
		\Big\{
			\int_0^\infty \frac{\alpha^2~\de \alpha}{1+\exp[\beta(\sqrt{\alpha^2+\gamma^2}-\sqrt{N_c}\frac{\mu}{\mu_0})]}
			-(\star)
		\Big\}
		\;.
	\label{eq:tzero_curves}
\qe
\begin{figure}[htb]
\centerline{
	\includegraphics[width=0.5\textwidth]{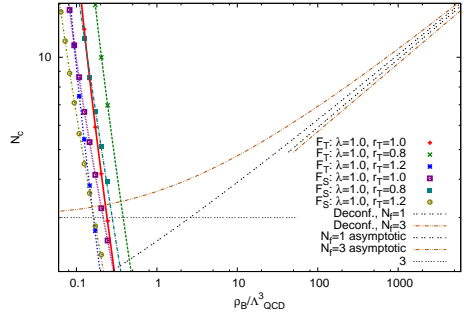}}
	\caption{
		Log-log plot of the $T=0$ deconfinement curves in the $\rho$-$N_c$ plane from Eq.~\ref{eq:tzero_curves}
		(and their large-$N_c$ limit) for $N_f=1,3$,
		compared to the percolation curves obtained as described in Section \ref{sec:percmodel} for the
		propagators $F_T$ and $F_S$ and some representative choices of the associated parameters.
		The horizontal line marks $N_c=3$.
	}
\label{Fig:tzero_curves}
\end{figure}
This can be solved exactly at zero temperature ($\beta\to\infty$), as shown in Fig.~\ref{Fig:tzero_curves} for $N_f=1,3$.
Contrary to the findings in \cite{percprl}, we see that a percolating confined phase seems possible also in our three-colour
world, but one has to push the density to extremely high values:
this, however, has to be taken with a grain of salt, the parametric assumptions made here being somewhat questionable
towards $N_c=3$.

\begin{figure}[htb]
\centerline{
	\includegraphics[width=0.4\textwidth]{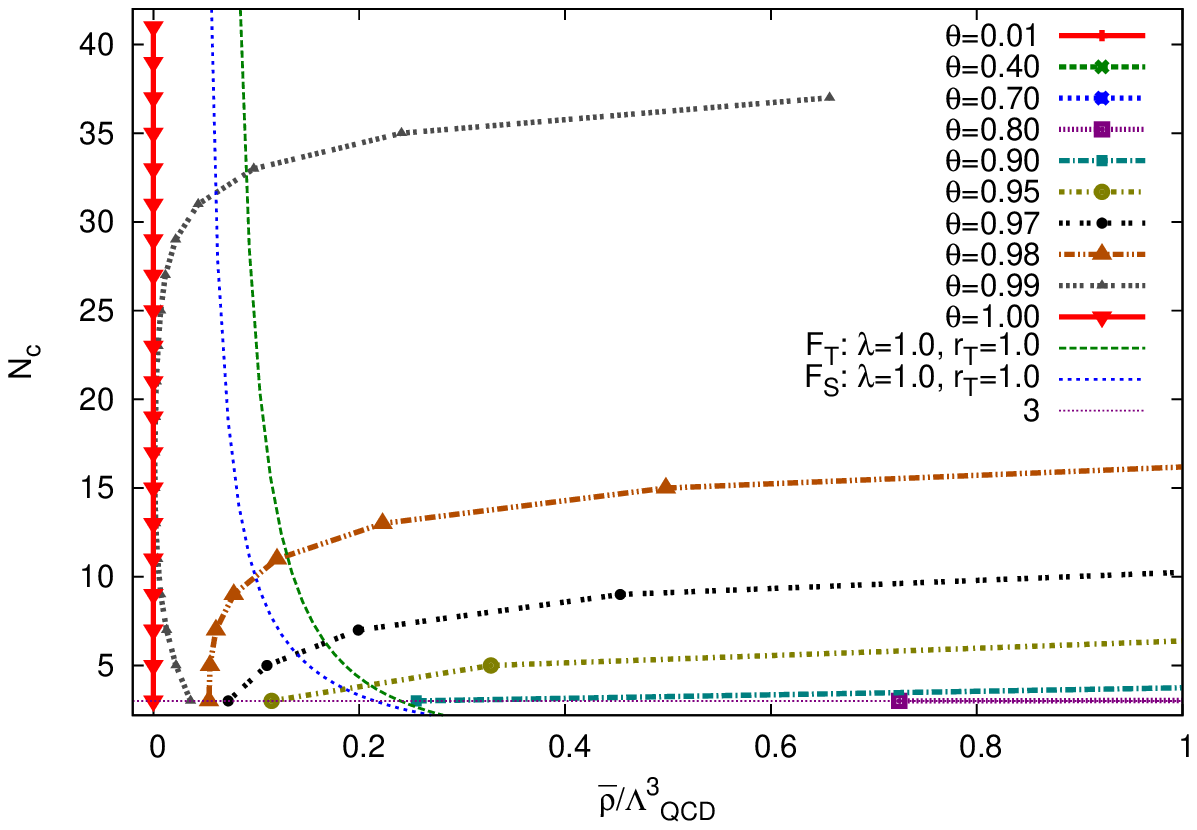}
	\includegraphics[width=0.4\textwidth]{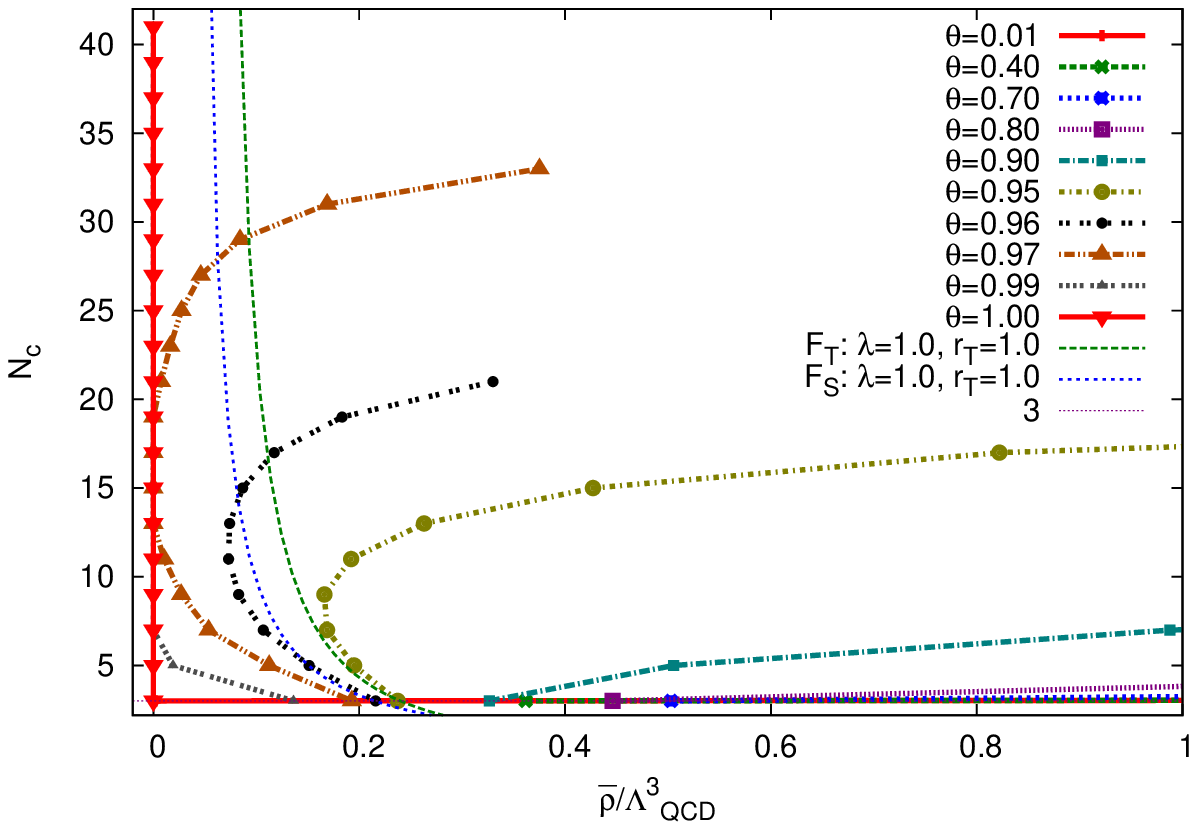}}
	\caption{Density curves at deconfinement in the $\rho$-$N_c$ plane, for various temperatures $T=\theta T_c$
		(datapoints; the lines are only a guide to the eye); the decreasing lines with no datapoints are
		parametrisations of two reference percolation curves for the two propagators $F_T$ and $F_S$.
		to the pure mesonic contribution.
		Plots on the left (right) are for $N_f=1$ ($3$).
	}
\label{Fig:densities}
\end{figure}

In order to extend the calculation to nonzero temperature, we will proceed numerically: here, there will be an antibaryonic contribution
and, most important, we will not neglect the presence of higher-spin baryons.
So, besides trading $\beta$ for $\theta$ using Eq.~\ref{eq:theta_param}, we consider that at a given $N_c=2Q+1$
we will have baryonic spins $(\eta+1)/2$, with $\eta=0,1,\ldots Q$, associated to an additional energy
$\eta \lqcd / N_c$ and with a spin-degeneracy of $2\eta+2$: we then replace, in Eq.~\ref{eq:tzero_curves},
$g_s \to \sum_\eta (2\eta+2)$ and add the cost of the spin-flip to the exponent in the denominator.
The resulting curves in the $\overline{\rho}$-$N_c$ plane are shown in Fig.~\ref{Fig:densities}.

Looking at Figs.~\ref{Fig:tzero_curves} and \ref{Fig:densities}, it is clear that in the $T$-$\mu$-$N_c$ space there is,
besides the usual confined and deconfined phases, also a ``confined but percolating'' regime (upper right region in the plots).
The question whether
it actually extends down to $N_c=3$ is obviously delicate (the roughness of this model introduces systematics uncertainties,
and it is not entirely clear if our world should be treated as a $N_f=3$ or $N_f=2$ setting, for instance),
but if there is the possibility, it certainly has to be investigated experimentally at densities at least several times that
of ordinary baryonic matter and at as low a temperature as possible.
An important cross-check, at least ideally given the massive numerical effort involved, would be to perform lattice experiments
at high $N_c$.

\section{Conclusions}
The percolation model presented here, although rather qualitative, shows the possible presence of a phase characterised by
confinement and, simultaneously, large-scale correlations: in \cite{percprl} it was argued that this could be identified
with the ``quarkyonic'' phase proposed in \cite{quarkyonic}.
In this work we examine the shape, in the $T$-$\mu$-$N_c$ space, of the region where this regime occurs,
finding that, provided the densities are high enough, there is a substantial possibility to observe its features
in our world too.
Based on the analogy between the percolating phase in hadronic matter and the metal-insulator transition, possible
experimental signatures for this phase can be suggested \cite{percdensity}.
A caveat is that, since the model is parametrically constructed in the large-$N_c$ spirit, its validity at $N_c=3$ could be hampered
by deviations from the expected asymptotic scaling:
at $N_c=3$ nuclear matter is a glass rather than a crystal, and nucleons are not fully classical objects.

\end{document}